\documentclass[a4paper, 10pt]{report}

\usepackage[cp1250]{inputenc}
\usepackage[T1]{fontenc}
\usepackage{amsfonts}
\usepackage{graphicx}
\usepackage{amsmath}

\usepackage{amssymb}

\usepackage{fancyheadings}
\usepackage{bibtopic}
\usepackage{color}
\usepackage{ulem}

\AtBeginDocument{}

\begin{document}

{\LARGE \bf

\centerline{Effect of radiation-like solid}

\centerline{on {small-scale CMB anisotropies}}

}

\vskip 1cm
\begin{center}
{Peter Mészáros and Vladimír Balek}

\vskip 2mm {\it Department of Theoretical Physics, Comenius
University, Bratislava, Slovakia}

\vskip 4mm July 12, 2017 \vskip 1cm {\bf Abstract}
\end{center}

\vskip 5mm We compute the CMB angular power spectrum in the
presence of {a} radiation-like solid {-- elastic matter} with the
same pressure to energy density ratio as radiation but with
nonzero shear modulus. {For the values of shear modulus that are
close enough to zero, so that the effect of the solid on
large-scale anisotropies remains within cosmic variance, we find
that there is an observable effect of the solid on small-scale
anisotropies.}

\vskip 1.5cm
{\bf{\Large 1 Introduction}}

\vskip 5mm The idea of solid matter was {introduced into cosmology
in} an attempt to give an alternative explanation of the
acceleration of the universe, see \cite{1} where the dark energy was
replaced by a solid with negative pressure to energy density ratio
$w$. Further development of {the} theory \cite{2,3,4,5,6,7,8} included
works where inflation was driven by a solid (solid inflation)
\cite{9,10,11,12,13}. An important example of {how such solid can
materialize} are cosmic strings and domain walls \cite{2,3,8}.

\vskip 2mm The effect of a solid can be obtained by any mechanism
leading to {nondiagonal 3-space components of energy-momentum
tensor of the desired form.} {In particular,} one can {replace
internal coordinates of the solid by three scalar fields
$\phi^A(x)$ and consider a Lagrangian $\mathcal{L}[\phi]$
depending on them \cite{10,11}.}

\vskip 2mm To extend the parameter space of the theory one can
consider also a solid with positive pressure to energy density
ratio \cite{14}. An important special case is the {\it
radiation-like solid} with $w = 1/3$, which does not {influence}
the evolution of the unperturbed universe if the total energy
density of radiation and solid equals the energy density of
radiation in the standard case. Materialization of such a solid
could be a Coulomb crystal with relativistic Fermi gas of moving
particles, or a network of speculative 'spring-like' strings with
energy inversely proportional to their length.

\vskip 2mm {Radiation-like solid, like any matter that has $w \ge
-1/3$, produces decelerating expansion of the universe. Thus, it
cannot be regarded as an alternative to dark energy. The reason
why we are studying it is that it would, if present in the
universe, modify the evolution of perturbations and leave an
imprint on CMB anisotropies.}

\vskip 2mm {The solid component of matter considered in cosmology
differs greatly from the solids known from everyday experience,
since it has to be stretched by many orders of magnitude in the
course of expansion of the universe, and still remain solid. Such
behavior is necessary if we want to preserve the cosmological
principle. If the solid was motionless while the other components
of cosmic matter were expanding, different observers would see it
in different states of motion, hence the universe would look
differently for them.}

\vskip 2mm Solid matter affects the evolution of perturbations if
it appears with flat internal geo\-metry and nonzero shear stress
acting in it. Such solidification cannot occur in pure radiation
and must be related to some other kind of particles distributed
anisotropically before solidification. {The particles could
possibly appear} in the universe as a remnant of solid inflation.

\vskip 2mm However {well the current observations are} explained
by standard theory with radiation, baryonic matter, cold dark
matter and dark energy, future observations could {be nevertheless
lacking in full explanation within} the standard theory and some
nonstandard theories, including the model presented in this work,
{might} become relevant.

\vskip 2mm The effect of {a} radiation-like solid on
long-wavelength perturbations and {on the} coefficients of angular
power spectrum of cosmic microwave background (CMB) {with} very
low multipole moments was studied in \cite{15}. {In the
long-wavelength limit} the sound speed does not appear in the
equations for perturbations (see (7.69) in \cite{16}), {therefore}
one can suppose the cold dark matter (CDM) to be coupled to the
baryon-radiation plasma. Also, considering neutrinos {to be}
coupled to photons {should} not change {the} results
qualitatively.

\vskip 2mm In this work we study {how a} radiation-like solid
appearing {in the universe} shortly after inflation {influences}
the CMB angular power spectrum {for arbitrary} multipole moments.
{To get the high moments right, one must consider} a model in
which the CDM and neutrinos are decoupled from the
baryon-radiation plasma. {Rather than} merely extending the
existing theory {of} a universe filled with ideal fluid, {which
uses gauge invariant variables as basic objects} \cite{W,16}, we
build the theory from the scratch in the proper-time comoving
gauge \cite{17}. In {section 2} we derive equations governing the
evolution of perturbations in this gauge, in {section 3} we
enumerate important effects contributing to the CMB anisotropies,
in section 4 we summarize the effects of the radiation-like solid
on the CMB anisotropies and in the last section we discuss
results. {To complete the analysis, in the appendix we solve}
equations {from section 2 analytically} in the long-wavelength
limit. We use the signature {of the metric tensor} $(+ - - -)$ and
{the} units in which $c = 16\pi G = 1$.

\vskip 1.5cm {\bf{\Large 2 Equations for perturbations}}

\vskip 5mm Observations of the CMB anisotropies can be explained
by the perturbation theory and of the three kinds of
perturbations, scalar, vector and tensor, scalar perturbations
appear to {give} the dominant contribution to the observed
anisotropies. We will consider scalar perturbation theory in a
flat Friedmann--Robertson--Walker--Lemaître (FRWL) universe filled
with radiation, baryonic matter and CDM, which appears to be a
realistic model of the universe in the pre-recombination era.
{H}owever, we will add {an} elastic radiation-like matter with
nonzero shear modulus to the other matter components. The presence
of the radiation-like solid{, which has} the pressure to energy
density ratio $w = 1/3$ {just like radiation (photons and, in the
early universe, neutrinos)}, does not change the evolution of the
unperturbed universe {provided that} the total energy density of
all radiation-like components is unchanged.

\vskip 2mm We will use {\it proper-time comoving gauge} \cite{17} in
which the $(00)$-component of the metric tensor is unperturbed and
the shift vector $\delta\mathbf{x}$ is zero for radiation-like
solid as well as for all {kinds of} matter coupled {to} it. In
this gauge the scalar part of metric takes the form
\begin{eqnarray}
ds^{(S)2} = a^2{\left\{d\eta^2 + 2B_{,i} d\eta dx^i - [(1 - 2\psi)
\delta_{ij} - 2E_{,ij}]dx^i dx^j\right\}}, \label{eq:01}
\end{eqnarray}
where $a$ {is} the scale parameter, $\eta$ is the conformal time
and $B$, $\psi$ and $E$ are the functions describing scalar
perturbations. If the {only matter coupled to the radiation-like
solid is the baryon-radiation plasma}, the scalar part of
energy-momentum tensor is
\begin{eqnarray}
& & T_0^{\phantom{0}0} = \rho + {\rho_{+}^{bs\gamma}} (3\psi +
\mathcal{E}) + \delta\rho^d + \delta\rho^\nu, \qquad
T_i^{\phantom{i}0(S)} = {\rho_{+}^{bs\gamma}}B_{,i} +
\rho^d\vartheta_{,i} + \rho_{+}^\nu\sigma_{,i},
\nonumber\\
& & T_i^{\phantom{i}j(S)} = -p\delta_i^j -
K(3\psi+\mathcal{E})\delta_i^j - 2\mu E^T_{,ij} -
2\mathcal{N}a^{-1}E^T_{,{\eta ij}}{-
(1/3)\delta\rho^\nu\delta_i^j}, \label{eq:02}
\end{eqnarray}
where $\rho$ is the total energy density, {$\rho_{+} = \rho + p$,
$p$ being pressure}, $\rho^d$ is the energy density of the CDM,
$\rho^\nu$ is the energy density of neutrinos, ${\rho^{bs\gamma}}$
is the energy density of {the system of} baryons, solid and
{photons}, $\delta\rho^d$ and $\delta\rho^\nu$ are perturbations
of the {energy density of CDM and neutrinos}, $\vartheta_{,i}$ and
$\sigma_{,i}$ are scalar parts (longitudinal parts in Helmholtz
decomposition) of the velocity of CDM and neutrinos with respect
to other matter, $K{=(1/3)\rho_+^{s\gamma}}$ is the compressional
modulus of the {system of solid and photons}, $\mu$ is the shear
modulus of the radiation-like solid, $\mathcal{N}$ is the shear
viscosity coefficient due to the coupling of photons and baryonic
matter before recombination (which is responsible for {\it Silk
damping}), $\mathcal{E}=\triangle E$ and $E^T_{,ij}\equiv
E_{,ij}-(1/3)\triangle E \delta_{ij}$ is {the} traceless part of
the tensor $E_{,ij}$.

\vskip 2mm Energy-momentum tensor with nondiagonal spatial
components can be obtained from {any given Lagrangian}
$\mathcal{L}[\phi]$ which {is} invariant under internal rotations
and translations \cite{10,11},
\begin{eqnarray}
& & \phi^A \to M^A_{\phantom{A}B}\phi^B, \qquad \phi^A \to \phi^A
+ C^A, \nonumber\\
& & M^A_{\phantom{A}B} \in SO(3), \qquad C^A\in\mathbb{R}^3,
\qquad A{,B}=1,2,3, \label{eq:03}
\end{eqnarray}
where the capital indices are raised and lowered by the Euclidean
metric $\delta_{AB}$. If the object described by the theory is
solid matter, the three-component field $\phi^A$ is interpreted as
the so-called {\it spatial internal coordinates} which move with
the matter. {As a result,} in the perturbed FRWL universe we have
\begin{eqnarray}
\phi^A(x) = \delta^A_i x^i - \delta x^A(x),
\label{eq:04}
\end{eqnarray}
{where} the fields $\delta x^A$ {describing} the perturbed state
of the solid play the role of Goldstone boson fields in {the field
formulation of the} theory. The Lagrangian leading to the
energy-momentum tensor (\ref{eq:02}) {up} to the first order {of}
perturbation theory, except for the Silk damping term, is \cite{14}
\begin{eqnarray}
\mathcal{L} = \mathcal{L}_0 + c_1 \partial_\mu \phi^A \partial^\mu
\phi_A +(c_2 \delta_{AB}\delta_{CD} + c_3 \delta_{AC}\delta_{BD})
\partial_\mu \phi^A \partial^\mu \phi^B \partial_\nu \phi^C \partial^\nu \phi^D,
\label{eq:05}
\end{eqnarray}
where $\mathcal{L}_0=\rho+9(K-\rho_+)/8$, $c_1=(3K-\rho_+)a^2/4$,
$c_2=({\lambda}+\rho_+)a^4/8${, $\lambda$ being the first Lame
coefficient related to compressional modulus by the formula $K =
\lambda + 2\mu/3$, and} $c_3=(\mu-\rho_+)a^4/4$. The Silk damping
term is {not obtained from the Lagrangian, but computed directly}
from the expression of energy-momentum tensor for an imperfect
fluid \cite{16}.

\vskip 2mm For the variables $\psi$, $\mathcal{E}$, $B$,
{$\vartheta$, $\delta\rho^d$, $\sigma$ and $\delta\rho^\nu$} we
have equations
\begin{eqnarray}
& & 4\psi'= a^2({\rho_{+}^{bs\gamma}}B+\rho^d \vartheta+\rho^\nu
\sigma),
\label{eq:06}\\
& & 4\triangle(\mathcal{H}B+\psi)-4\mathcal{H}(3\psi+\mathcal{E})'
=
a^2({\rho_{+}^{bs\gamma}}(3\psi+\mathcal{E})+\delta\rho^d+\delta\rho^\nu),
\label{eq:07}\\
& & {\rho_{+}^{bs\gamma}}B'=\mathcal{H}(3K-{\rho_{+}^{bs\gamma}})B
+ 3K\psi+(K+4\mu/3)\mathcal{E} +4\mathcal{N}a^{-1}\mathcal{E}'/3,
\label{eq:08}\\
& & \vartheta'=-\mathcal{H}\vartheta,
\label{eq:09}\\
& & {(\delta\rho^d/\rho^d)'-(3\psi+\mathcal{E})'=\triangle
(\vartheta-B),}
\label{eq:10}\\
& & {4\sigma'=\delta\rho^\nu/\rho^\nu,
\label{eq:11}}\\
& & 3(\delta\rho^\nu/\rho^\nu)'-4(3\psi+\mathcal{E})'=4\triangle
(\vartheta-B), \label{eq:12}
\end{eqnarray}
where the prime denotes differentiation with respect to $\eta$ and
$\mathcal{H}=a'/a$. Equations (\ref{eq:06}) and (\ref{eq:07}) are
derived from {$\binom 0i$ and $\binom 00$} components of Einstein
field equations, $2G_{i}^{\phantom{i}0}=T_{i}^{\phantom{i}0}$ and
$2G_{0}^{\phantom{0}0}=T_{0}^{\phantom{0}0}$, equations
(\ref{eq:08}), (\ref{eq:09}) and (\ref{eq:11}) are obtained from the
{momentum} conservation law,
$T_{i\phantom{\mu};\mu}^{\phantom{i}\mu}=0$, {written for the
system of baryon-radiation plasma and radiation-like solid}, CDM
and neutrinos respectively, and equations (\ref{eq:10}) and
(\ref{eq:12}) follow from the energy conservation law,
$T_{0\phantom{\mu};\mu}^{\phantom{0}\mu}=0$, {written} for CDM and
neutrinos. All equations are obtained in the first order of the
perturbation theory in the case when the internal geometry of the
radiation-like solid is flat.

\vskip 2mm Equations (\ref{eq:06}){--}(\ref{eq:12}) are not
invariant under coordinate transformations because the proper-time
comoving gauge allows for a residual transformation
$\eta\to\eta+a^{-1}\delta t(\mathbf{x})$, where $\delta
t(\mathbf{x})$ is {the} local shift of the moment at which the
time count starts. The function $\mathcal{E}$ is invariant under
this {transformation} and $B$, $\psi$, CDM energy density contrast
$\delta^d=\delta\rho^d/\rho^d$, neutrino energy density contrast
$\delta^\nu=\delta\rho^\nu/\rho^\nu$, $\vartheta$ and $\sigma$ can
be {rewritten as
\begin{eqnarray}\label{eq:8}
& & B = \mathcal{B}+\chi, \qquad \psi=-\mathcal{H}\chi, \nonumber\\
& & \delta^d={\hat\delta^d}-3\mathcal{H}\chi, \qquad
\vartheta=\hat{\vartheta}+\chi, \nonumber \\
{} & & \delta^\nu={\hat\delta^\nu}-4\mathcal{H}\chi, \qquad
\sigma=\hat{\sigma}+\chi,
\end{eqnarray}
where $\chi$ transforms as $\chi\to\chi+a^{-1}\delta
t(\mathbf{x})$ and $\mathcal{B}$, $\hat{\delta}^d$,
$\hat{\delta}^\nu$, $\hat{\vartheta}$ and $\hat{\sigma}$ are
invariant. It is convenient to introduce a rescaled time
$\zeta=\eta/\eta_*$, where $\eta_*=\eta_{eq}/(\sqrt{2}-1)$ and
$\eta_{eq}$ denotes the moment when the energy density of matter
is equal to the energy density of radiation. {As long as} the
contribution of dark energy to the total energy density is
negligible, which is {surely} the case before recombination, the
scale parameter can be {written} as $a=a_{eq}\zeta(\zeta+2)$.

\vskip 2mm Using (\ref{eq:06}) and (\ref{eq:8}) and considering
perturbations of the form of a plane wave with the comoving wave
vector $\mathbf{k}$, equations (\ref{eq:07}){--}(\ref{eq:12}) can
be rewritten into the invariant form
\begin{eqnarray}
& & \mathcal{E}'=-(s^2+3{\alpha_{bs\gamma}}\tilde{\mathcal{H}}^2)
\tilde{\mathcal{B}}
-{\alpha_{bs\gamma}}\tilde{\mathcal{H}}\mathcal{E} -
3\alpha_d\tilde{\mathcal{H}}^2\Theta
-\alpha_d\tilde{\mathcal{H}}\hat{\delta}^d -
3\alpha_\nu\tilde{\mathcal{H}}^2\Sigma
-\frac{3}{4}\alpha_\nu\tilde{\mathcal{H}}\hat{\delta}^\nu,
\label{eq:14}\\
& & \tilde{\mathcal{B}}'=(3c_{s0}^2+{\alpha_{bs\gamma}}-1)
\tilde{\mathcal{H}}\tilde{\mathcal{B}}+
c_{s||}^2\mathcal{E}+\Xi\mathcal{E}'+\alpha_d\tilde{\mathcal{H}}
\Theta+\alpha_\nu\tilde{\mathcal{H}}\Sigma,
\label{eq:15}\\
& &
\Theta'={\alpha_{bs\gamma}}\tilde{\mathcal{H}}\tilde{\mathcal{B}}
+(\alpha_d-1)\tilde{\mathcal{H}}\Theta
+\alpha_\nu\tilde{\mathcal{H}}\Sigma,
\label{eq:16}\\
& &
{\hat{\delta}^d}{}'=\mathcal{E}'+s^2(\tilde{\mathcal{B}}-\Theta),
\label{eq:17}\\
& &
\Sigma'={\alpha_{bs\gamma}}\tilde{\mathcal{H}}\tilde{\mathcal{B}}
+\alpha_d\tilde{\mathcal{H}}\Theta
+\alpha_\nu\tilde{\mathcal{H}}\Sigma+\frac{1}{4}\hat{\delta}^\nu,
\label{eq:18}\\
& &
{\hat{\delta}^\nu}{}'=\frac{4}{3}\mathcal{E}'+\frac{4}{3}s^2
(\tilde{\mathcal{B}}-\Sigma),
\label{eq:19}
\end{eqnarray}
where the prime denotes differentiation with respect to $\zeta$,
{$s=k\eta_{*}$, $\tilde{\mathcal{H}}=a'/a$ with the redefined
prime,} $\tilde{\mathcal{B}}=\mathcal{B}/\eta_{*}$,
$\Theta=\hat{\vartheta}/\eta_{*}$, $\Sigma=\hat{\sigma}/\eta_{*}$,
${\alpha_{bs\gamma}}=3{\rho_{+}^{bs\gamma}}/(2\rho)$,
$\alpha_d=3\rho^d/(2\rho)$, $\alpha_\nu=3\rho_{+}^\nu/(2\rho)$
and $\Xi=4\mathcal{N}/(3a{\rho_{+}^{bs\gamma}}\eta_{*})$. The
two sound speeds appearing in equation (\ref{eq:15}), auxiliary
sound speed {$c_{s0}$ and} longitudinal sound speed of the
baryon-radiation plasma with radiation-like solid coupled to it
{$c_{s||}$}, are defined as
\begin{eqnarray}\label{eq:20}
c_{s0}^2=\frac K{{\rho_{+}^{bs\gamma}}}, \qquad
c_{s||}^2=c_{s0}^2+\frac{4}{3}\frac{\mu}{\rho_{+}^{bs\gamma}}=(1+3\xi)c_{s0}^2,
\end{eqnarray}
where $\xi$ is the dimensionless shear modulus defined as
\begin{eqnarray}\label{eq:21}
\xi{=}\frac{\mu}{{\rho^{s\gamma}}},
\end{eqnarray}
{$\rho^{s\gamma}$ being the} sum of energy densities of {solid and
photons}. The shear viscosity coefficient $\mathcal{N}$ entering
equation (\ref{eq:15}) is proportional to the mean free time for
photon scattering, which is {in turn inversely} proportional to
electron ionization fraction. {In our computation, we took into
account that this fraction drops smoothly from 1 to a value close
to zero during recombination} (see \S 3.6.2 and \S 3.6.3 in
\cite{16}).

\vskip 2mm Equations (\ref{eq:14}){--}(\ref{eq:19}) {describe
completely the evolution of perturbations in the pre\-sence of
radiation-like solid. They generalize} equations (4) in \cite{15}
valid in long-wavelength limit {only}.

\vskip 2mm Perturbations generated {during} inflation {are}
long-wavelength (superhorizon) {afterwards}, therefore the initial
conditions for equations {describing their evolution}, {set at
some moment $\zeta_{inf}$ shortly after the end of inflation}, are
$\mathcal{E}(\zeta_{inf})=(-9/2)\Phi^{(0)}$,
$\tilde{\mathcal{B}}(\zeta_{inf})=0$, $\Theta(\zeta_{inf})=0$,
$\hat{\delta}^d(\zeta_{inf})=(-9/2)\Phi^{(0)}$,
$\Sigma(\zeta_{inf})=0$,
$\hat{\delta}^\nu(\zeta_{inf})=-6\Phi^{(0)}$, where $\Phi^{(0)}$
is the primordial Newtonian potential. {Suppose} the time of
solidification $\zeta_s$ {(the time when the radiation-like solid
was formed) was} greater than {the} time $\zeta_{inf}$ {but} much
smaller than {the time} $\zeta_{eq}$ ({the} time of
matter--radiation equality). {Then,} the initial conditions {at
some moment $\zeta_{in} > \zeta_s$,} when the numerical
{integration} of the equations for perturbations starts, {can be
obtained by solving the equations between the times $\zeta_{inf}$
and $\zeta_{in}$ analytically,} neglecting {the} contribution of
the CDM and baryonic matter to the energy density and considering
the primordial perturbations to be superhorizon up to the moment
$\zeta_{in}$. In this way we find
\begin{eqnarray}
& & \mathcal{E}(\zeta_{in})=
-\frac{9}{4n}\left(M\hat{\zeta}^{-m}-m\hat{\zeta}^{-M}\right),
\nonumber\\
& & \tilde{\mathcal{B}}(\zeta_{in})=
\frac{9\zeta_s}{24n}\left[M(M+1)\hat{\zeta}^M-m(m+1)\hat{\zeta}^m\right],
\nonumber\\
& & \Theta(\zeta_{in})=
\tilde{\mathcal{B}}(\zeta_{in})+\frac{9}{4n}\xi\zeta_s\left(\hat{\zeta}^M-\hat{\zeta}^m\right),
\nonumber\\
& & \hat{\delta}^d(\zeta_{in})=\mathcal{E}(\zeta_{in}),
\nonumber\\
& & \Sigma(\zeta_{in})=
\tilde{\mathcal{B}}(\zeta_{in})-\frac{9}{2}\xi\zeta_{in},
\nonumber\\
& & \hat{\delta}^\nu(\zeta_{in})=
\frac{4}{3}\mathcal{E}(\zeta_{in})-36\xi, \label{eq:22}
\end{eqnarray}
where $\hat{\zeta}=\zeta_{in}/\zeta_s$, $n=\sqrt{1-24\xi}/2$,
$m=1/2-n$ and $M=1/2+n$.

\vskip 2mm {We have assumed that the moment $\zeta_{inf}$ occurred
shortly, but not immediately, after the end of inflation, so that
the decaying part of perturbations was negligible then. We have
also assumed that the moment $\zeta_{in}$ followed soon enough
after the moment $\zeta_s$, so that it occurred much sooner than
the time of equality $\zeta_{eq}$ and the short-scale
perturbations were superhorizon then. The latter assumption at the
same time enabled us to use the law $a \propto \zeta$ valid for
radiation-dominated universe, when deriving the conditions
(\ref{eq:22}). The constraints on $\zeta_{inf}$ and $\zeta_{in}$
leave us with a wide interval of possible values of the
solidification time $\zeta_s$. It turns out, however, that the
results practically do not depend on the actual value of
$\zeta_s$, so that we could have chosen $\zeta_s = 10^{-9}$, which
corresponds to energy about 500 MeV, in our calculations, although
the solid presumably appeared at an energy scale much greater than
the energies accessible at present-day accelerators.

\vskip 2mm The perturbations were supposed to be purely adiabatic
at the moment $\zeta_{inf}$. If entropic perturbations emerged
from inflation, the time $\zeta_{inf}$ should be shifted to the
first period of local thermal equilibrium with no non-zero
conserved quantities, see \S 5.4 in \cite{W}. Note that after
solidification, the perturbations do not retain the simple form
derived in \cite{W} for adiabatic perturbations. For an analysis
of this effect, see \S 9 in \cite{10}.}

\vskip 2mm Instead of the proper-time comoving gauge, one often
uses {\it Newtonian gauge} in which the scalar part of space-time
metric is diagonal. Metric in the Newtonian gauge {is given} by
two potentials invariant under coordinate transformations, $\Phi$,
called {\it Newtonian potential}, and $\Psi$, as
$g_{00}=a^2(1+2\Phi)$ and $g_{ij}=-a^2(1-2\Psi)\delta_{ij}$. The
functions $\Phi$ and $\Psi$ can be {written} as \cite{16}
\begin{eqnarray}
\Psi=\mathcal{H}(\mathcal{B}-E'), \qquad \Phi=\Psi-\mu a^2E,
\label{eq:23}
\end{eqnarray}
where the difference between $\Phi$ and $\Psi$ is given by the
traceless part of the equation
$2G_{i}^{\phantom{i}j}=T_{i}^{\phantom{i}j}$. The Silk damping is
omitted here since its contribution to the second equation in
(\ref{eq:23}) is negligible.

\vskip 2mm Characteristics of matter, density contrast $\delta =
\delta \rho/\rho$ and ``velocity potential'' $\phi$ defining the
scalar part of velocity according to $u^{(S)}_i = \phi_{,i}$, can
be computed in Newtonian gauge as well. In this way we obtain
invariant functions $\bar \delta$ and $\bar \phi$, which are often
viewed as {\it physical} quantities. For CDM and neutrinos, the
physical density contrasts and ''velocity potentials'' are
\begin{eqnarray}
\bar{\delta}^d=\hat{\delta}^d+3\mathcal{H}(\mathcal{B}-E'), \qquad
\bar{\vartheta}=\hat{\vartheta}+\mathcal{B}-E', \nonumber\\
\bar{\delta}^\nu=\hat{\delta}^\nu+4\mathcal{H}(\mathcal{B}-E'),
\qquad \bar{\sigma}=\hat{\sigma}+\mathcal{B}-E'.
\label{eq:24}
\end{eqnarray}
Another useful relation,
\begin{eqnarray}
\overline{\delta{\rho}}^{bs\gamma}={\rho_{+}^{bs\gamma}}(3\Psi+\mathcal{E}),
\label{eq:25}
\end{eqnarray}
is valid also for separate components of the system consisting of
baryons, {solid and photons}.

\vskip 2mm Note that the velocity potentials $\hat{\vartheta}$ and
$\hat{\sigma}$ define the spatial part of 4-velocity with the
lower index. In order to acquire physical velocities one must rise
the index with respect to the perturbed metric to obtain the
potentials $\mathcal{B}-\hat{\vartheta}$ for the CDM physical
velocity and $\mathcal{B}-\hat{\sigma}$ for the neutrino physical
velocity, both with respect to the baryon-radiation plasma.
Furthermore{,} for the long-wave perturbations we have
$\overline{\delta}^d=(3/4)\overline{\delta}^\gamma$ and
$\overline{\delta}^\nu=\overline{\delta}^\gamma$ \cite{18}, and since
$\overline{\delta}^\gamma=4\Psi+4\mathcal{E}/3$ (see equation
(\ref{eq:25})) and $\overline{\delta}^d=\hat{\delta}^d+3\Psi$,
$\overline{\delta}^\nu=\hat{\delta}^\nu+4\Psi$ (see the first
equation in (\ref{eq:23})), we can see that the differences
$\hat{\delta}^d-\mathcal{E}$ and
$\hat{\delta}^\nu-(4/3)\mathcal{E}$ are deviations of the actual
density contrasts from their limit values.

\vskip 1.5cm {\Large {\bf 3 Fluctuations of the CMB temperature}}

\vskip 5mm The CMB anisotropies are given by perturbations around
the time of last scattering and effects influencing photons during
their propagation to the observer, {which can be described by the
relativistic Boltzmann equation for polarization-dependent
distribution function of photons. The solution for temperature
fluctuations $\delta T/T$ can be expressed in the form of
line-of-sight integral, and if we restrict ourselves to scalar
fluctuations, we can write it as a sum of two terms (see \S 7.1 in
\cite{W}),
\begin{eqnarray}
\frac{\delta T}{T}(\eta_0,\mathbf{l})=\left(\frac{\delta T}{T}
\right)_\textrm{early}(\eta_0,\mathbf{l})+\left(\frac{\delta
T}{T}\right)_\textrm{ISW}(\eta_0,\mathbf{l}), \label{eq:26}
\end{eqnarray}
where $\eta_0$ is the conformal time today and $\mathbf{l}$ is the
unit vector pointing towards the observer from the place in the
sky from which the radiation is coming. The first term represents
fluctuations which appeared when there was significant amount of
free electrons in the universe (this includes the period of
reionization, which is however not considered here), and the
second term, called {\it integrated Sachs--Wolfe effect},
represents fluctuations which arise from the action of metric
perturbations on radiation during its propagation to the observer.

\vskip 2mm The early part of temperature fluctuations simplifies
if we assume that photons are in local thermal equilibrium during
the whole period of recombination, and that they can scatter for
the last time at any moment $\eta_l$ with the probability
distribution given by their mean free time at that moment. The
line-of-sight integral then reduces to an integral over the times
$\eta_l$, with the integrand that can be written as
\begin{eqnarray} \left(\frac{\delta
T}{T}\right)_\textrm{early}(\eta_0,\mathbf{l};\eta_l)
=\frac{1}{4}\delta_\gamma(\eta_l,\mathbf{l})+\Phi(\eta_l,\mathbf{l})+\mathbf{v}
(\eta_l,\mathbf{l})\cdot\mathbf{l}, \label{eq:27}
\end{eqnarray}
}where $\delta_\gamma=\overline{\delta\rho}^\gamma/\rho^\gamma$ is
the photon density contrast in Newtonian gauge (for simplicity, we
skip the bar over $\delta$), $\mathbf{v}$ is the local velocity of
the radiating matter, and $\delta_\gamma(\eta,\mathbf{l})$,
$\Phi(\eta,\mathbf{l})$ and $\mathbf{v}(\eta,\mathbf{l})$ denote
the values of functions $\delta_\gamma$, $\Phi$ and $\mathbf{v}$
at the conformal time $\eta$ and at the position where the photon
arriving in the direction $\mathbf{l}$ was at that time. The first
term in (\ref{eq:27}) is the local contribution to $\delta T/T$ as
computed from the proportionality $\rho_\gamma\propto T^4$
(Stefan--Boltzmann law), {the second term is called {\it
Sachs--Wolfe effect} and comes from the action of metric
perturbations on radiation in the place of last scattering, and
the third term is due to {\it Doppler effect}, which contributes
to the temperature fluctuations because the radiating matter at
the time of last scattering is not static.} Note also that
equation (\ref{eq:25}), when written for photons, implies that
$\delta_\gamma$ can be expressed in terms of the perturbations of
metric as $\delta_\gamma=4(\Psi+\mathcal{E}/3)$.

\vskip 2mm {The ISW part of temperature fluctuations is known to
be negligible, except for fluctuations with greatest wavelengths,
in case the universe is filled with ideal fluid. A question arises
whether this is the case also when the universe contains a solid
component. The answer is apparently ``yes'', as can be seen most
easily if we estimate the integral of $\Phi'+ \Psi'$, which
appears in the part of the angular power spectrum coming from
$(\delta T/T)_{ISW}$, by keeping the leading terms in the
expansion of $\Phi'+ \Psi'$ in $s$ only.}

\vskip 2mm The velocity $\mathbf{v}$ can be obtained from the
energy conservation law for radiation, written in Newtonian gauge,
\begin{eqnarray}
3(\delta_\gamma-4\Psi)'+4\nabla\cdot\mathbf{v}=0.
\label{eq:28}
\end{eqnarray}
(See (7.110) in \cite{16}, where $\Phi$ is to be replaced by
$\Psi$ in order that the equation stays valid in the presence of
solid matter.) Considering perturbations of the form of the a
plane wave with the comoving wave vector $\mathbf{k}$, we find
{\begin{eqnarray} \mathbf{v}(\eta_l,\mathbf{l})\cdot\mathbf{l}=
i\frac{3}{4}\frac{\mathbf{k}\cdot\mathbf{l}}{k^2}
\frac{\partial}{\partial\eta} (\delta_{\gamma}-4\Psi)_{\eta_l},
\label{eq:29}
\end{eqnarray}
so that the contribution of photons which scattered at the last
time at the moment $\eta_l$ to the fluctuations of temperature
seen in the direction $\mathbf{l}$ is approximately
\begin{eqnarray}
\frac{\delta T}{T}(\eta_0,\mathbf{l};\eta_l)=\int
\frac{d^3k}{(2\pi)^{3/2}} \left( \alpha(\mathbf{k})
+\beta(\mathbf{k}){\partial_0} \right)_{\eta_l}
e^{i\mathbf{k}\cdot\mathbf{l}(\eta_l-\eta_0)},
\label{eq:30}
\end{eqnarray}
where $\partial_0 = \partial/\partial \eta_0$ and
$\alpha(\mathbf{k})$ and $\beta(\mathbf{k})$ are defined as
\begin{eqnarray}
\alpha(\mathbf{k})=\frac{1}{4}\delta_{\gamma\mathbf{k}}+\Phi_{\mathbf{k}},
\qquad
\beta(\mathbf{k})=-\frac{3}{4k^2}\frac{\partial}{\partial\eta}
\left(\delta_{\gamma\mathbf{k}}-4\Psi_\mathbf{k}\right).
\label{eq:31}
\end{eqnarray}
}The functions $\delta_{\gamma\mathbf{k}}$, $\Phi_{\mathbf{k}}$
and $\Psi_{\mathbf{k}}$ entering here are amplitudes of the
corresponding perturbations {in the form of} plane waves, which
depend on the conformal time $\eta$ only.

\vskip 2mm In what follows we suppose that the dimensionless shear
modulus $\xi$ defined in (\ref{eq:21}) is small in absolute value.
{The numerical solution of equations for perturbations revealed
that for $|\xi|\lesssim 5\cdot 10^{-4}$ it is sufficient to
consider the functions $\alpha$ and $\beta$ in the form of Taylor
expansion up to the first order in $\xi$,
\begin{eqnarray}
\alpha(\mathbf{k})=\alpha_0(\mathbf{k})+\xi\alpha_1(\mathbf{k}),
\qquad
\beta(\mathbf{k})=\beta_0(\mathbf{k})+\xi\beta_1(\mathbf{k}).
\label{eq:32}
\end{eqnarray}}

The functions $\alpha_0$, $\alpha_1$, $\beta_0$ and $\beta_1$ can
be computed at any moment $\eta$ by numerical solution of
equations (\ref{eq:14})--(\ref{eq:19}) with the initial conditions
(\ref{eq:22}). {The equations are accurate in the tight-coupling
regime until some moment before the time of recombination $\eta_r$
defined below, and in the limit of sharp transition from opacity
to transparency we can use them up to the moment $\eta_r$. The
behavior of the functions $\alpha_0$, $\alpha_1$, $\beta_0$ and
$\beta_1$ at that moment is shown in fig. 1.} In the calculations
we have used the values $\zeta_s=10^{-9}$ and
$\zeta_{in}=10^{-5}$, {and for the viscosity coefficient entering
equation (\ref{eq:15}) we have used the formula
\begin{eqnarray}
\mathcal{N}=\frac{4}{15}\frac{\rho^\gamma}{\sigma_T n_e X_e},
\label{eq:33}
\end{eqnarray}
where $\rho^\gamma$ is the photon energy density, $\sigma_T$ is
Thomson cross section, $n_e$ is the total number density of
electrons and $X_e$ is the ionization fraction. For the redshift
$z$ up to $1200$ we have approximated the ionization fraction by
the formula (3.202) in \cite{16}, and for $z$ over $1200$ we have
used an exponential function converging to $1$ for $z$ going to
infinity and matching the low-redshift function together with its
derivative at $z=1200$. Note that the formula (\ref{eq:33}) has
been derived without taking into account the polarization of
photons. In a more accurate formula, the factor $4/15$ is replaced
by $16/45$, see \S 6.4 in \cite{W}; however, the results presented
in the next section are not affected significantly by this
correction.}

\begin{figure}[htb]
\centering
\includegraphics[scale=0.15]{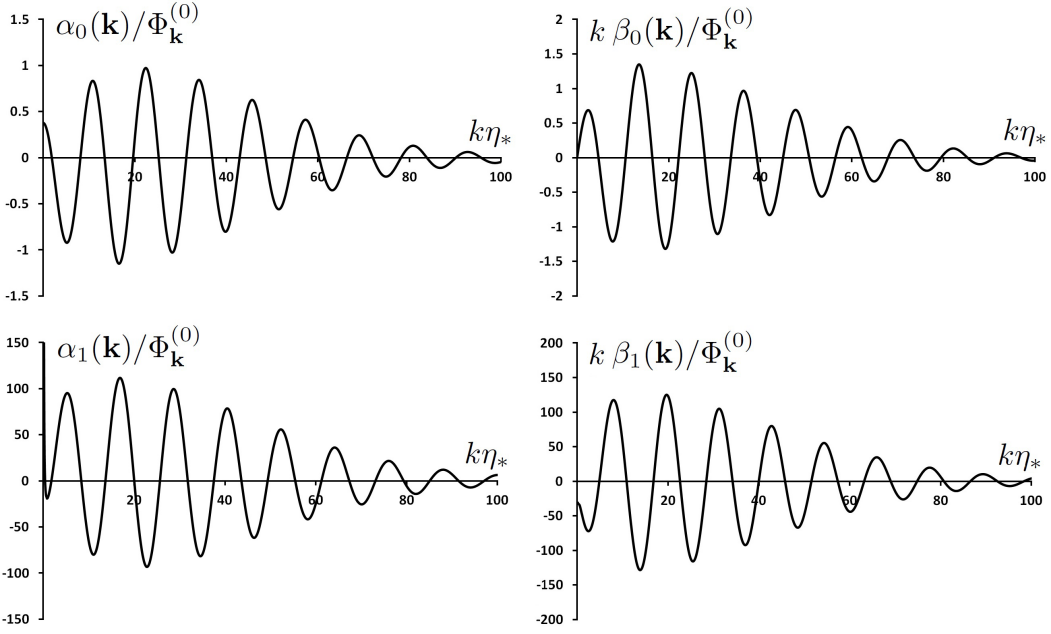}
\caption{Functions $\alpha_0$, $\alpha_1$, $k\beta_0$ and
$k\beta_1$ at the moment $\eta=\eta_r$, divided by the
primordial value of the Newtonian potential
$\Phi^{(0)}_{\mathbf{k}}$. The ratios depend only on the
comoving wavenumber $k$ and not on the direction of the vector
$\mathbf{k}$, because the equations governing the evolution of
perturbations are isotropic.}
\end{figure}

\vskip 2mm Note that $\alpha_1$ diverges in the long-wavelength
limit being proportional to $k^{-2}$ for small $k$. This
divergence occurs due to the gauge transformation in
(\ref{eq:23}), while variables in proper-time comoving gauge have
no divergences so that the perturbation theory remains applicable.
This is the conceptual reason for using proper-time comoving
gauge in our problem.

\vskip 2mm Due to the finite duration of recombination, the
observer sees photons which scattered for the last time at
different moments. This is called {\it finite thickness effect}.
The probability that the photon was scattered within the given
conformal time interval and then avoided further scattering
until the conformal time $\eta_0$, is given by the probability
density function
\begin{eqnarray}
\mathcal{P}(\eta)=\frac{d\mu(\eta)}{d\eta} e^{-\mu(\eta)}, \qquad
\mu(\eta)=\int\limits_{\eta}^{\eta_0} \frac{a}{\tau} d\eta,
\label{eq:34}
\end{eqnarray}
where $\tau$ is the mean free time for Thomson scattering, which
varies with time, and $\mu$ is {the} {\it optical depth}. {The
function} $\mathcal{P}$ is called {\it visibility function}. {Due
to the finite thickness effect,} the overall fluctuations of the
CMB temperature as seen by an observer at the present moment {can
be written as}
\begin{eqnarray}
\frac{\delta T}{T}(\eta_0,\mathbf{l}) =\int \frac{\delta
T}{T}(\eta_0,\mathbf{l};\eta_l) \mathcal{P}(\eta_l)d\eta_l.
\label{eq:35}
\end{eqnarray}
Since the visibility function can be approximated by a Gaussian
function (see (7.2.25) in \cite{W} or (9.55) in \cite{16}), the observed relative
anisotropies in the CMB temperature can be written as
\begin{eqnarray}
\frac{\delta T}{T}(\eta_0,\mathbf{l})=\int
\frac{d^3k}{(2\pi)^{3/2}} \left( \alpha(\mathbf{k})
+\beta(\mathbf{k}){\partial_0} \right)_{\eta_r}
e^{i\mathbf{k}\cdot\mathbf{l}(\eta_r-\eta_0)}
e^{-\Sigma k^2}, \label{eq:36}
\end{eqnarray}
where the conformal time of recombination $\eta_r$ is the time
when the visibility function is maximal and {the function $\Sigma$
is defined as} $\Sigma=(6\kappa^2\mathcal{H}(\eta_r)^2)^{-1}$,
$\kappa$ being the ratio of the ionization energy {in} the 2S
state of the hydrogen atom {to} the energy $k_B T_r$ corresponding
to the temperature of recombination $T_r$ (see \S 3.6.3 in
\cite{16}).

\vskip 1.5cm {\bf{\Large 4 Angular power spectrum}}

\vskip 5mm Relative anisotropies in the CMB temperature can be
expanded into spherical harmonics, with the coefficients that can
be written as scalar {products} of the anisotropy with {the
corresponding} spherical {harmonic,}
\begin{eqnarray}
\frac{\delta
T}{T}(\eta_0,\mathbf{l})=\sum_{lm}{a_{lm}Y_{lm}(\mathbf{l})},
\quad a_{lm}=\int d^2\mathbf{l}\frac{\delta
T}{T}(\eta_0,\mathbf{l})Y^{*}_{lm}(\mathbf{l}). \label{eq:37}
\end{eqnarray}
Since the CMB anisotropies are random, the coefficients $a_{lm}$
are random as well{, and} in order to compare the theory with the
{observational} data, correlation functions must be introduced.
The two-point correlation function is defined as
$C(\theta)=\left<\delta T (\eta_0,\mathbf{l}_1) \delta T
(\eta_0,\mathbf{l}_2){/T^2} \right>$, where $\theta$ is the angle
between $\mathbf{l}_1$ and $\mathbf{l}_2$ and the angular brackets
denote averaging over {observer's positions}. This function can be
rewritten as a sum over multipole moments
\begin{eqnarray}
C(\theta)=\frac{1}{4\pi}\sum_l{(2l+1) {C_l}P_l(\cos\theta)},
\label{eq:38}
\end{eqnarray}
where $C_l=\left<|a_{lm}|^2\right>$ are the coefficients of the
angular power spectrum of CMB and $P_l$ are Legendre polynomials.
The mean values $\left<|a_{lm}|^2\right>$ are independent on $m$
because there is no preferred direction in the universe, but for
the given observer there {always} exists statistical randomness
{in} the CMB anisotropies, so that the observed {values of}
$|a_{lm}|^2$ {with the given} multipole moment $l$ are not {the
same} for all $m$. The best estimate of the coefficients of
angular power spectrum is then given by averaging
\begin{eqnarray}
C_l = \frac{1}{2l+1}\sum_{m=-l}^l|{a_{lm}|^2},
\label{eq:39}
\end{eqnarray}
with an unavoidable error known as the {\it cosmic variance},
\begin{eqnarray}
\frac{\Delta C_l}{C_l} = \sqrt{\frac{2}{2l+1}}. \label{eq:40}
\end{eqnarray}

\vskip 5mm Using (\ref{eq:36}) and (\ref{eq:37}) we find
\begin{eqnarray}
C_l = \frac{2}{\pi} \int\limits_0^{\infty} \tau_l(k)^2 (2\pi)^3
\mathcal{P}_{\Phi}(k) k^2 dk, \label{eq:41}
\end{eqnarray}
where the functions $\tau_l(k)$ are defined as {\begin{eqnarray}
\tau_l(k) {=}
\left[\frac{\alpha(\mathbf{k})}{\Phi^{(0)}_{\mathbf{k}}}
j_l(k(\eta_0-\eta_r))+{\frac{k\beta(\mathbf{k})}{\Phi^{(0)}_{\mathbf{k}}}j_l'(k(\eta_0-\eta_r))}
\right] e^{-\Sigma k^2},
\end{eqnarray}
}$j_l$ being the spherical Bessel functions of the first kind, and
the function $\mathcal{P}_{\Phi}(k)$, known as {\it power
spectrum}, is defined by
\begin{eqnarray}
\left<\Phi^{(0)}_{\mathbf{k}}\Phi^{(0)}_{\mathbf{k}'}\right> =
(2\pi)^3
{\mathcal{P}_{\Phi}(k)\delta^{(3)}(\mathbf{k}+\mathbf{k}')}.
\label{eq:42}
\end{eqnarray}
The functions $\tau_l(k)$ depend only on the comoving wavenumber
$k$ because {they are solutions} to the equations governing the
evolution of perturbations, which are isotropic, with {the
primordial Newtonian potential $\Phi^{(0)}_{\mathbf{k}}$ appearing
in the initial conditions replaced by 1}. The power spectrum is
usually written as $\mathcal{P}_{\Phi}(k) \propto k^{n_s-4}$,
where $n_s$ is the {\it spectral index} given by {the}
inflationary {scenario describing the formation of perturbations.
In our calculations we have used the {\it Planck} value of
spectral index $n_s = 0.96$, which corresponds to the primordial
spectrum slightly tilted downwards.}

\vskip 2mm {Define} the variance of the {coefficients} of angular
power spectrum due to {the} presence of radiation-like solid as
\begin{eqnarray}
\frac{\Delta_\xi C_l}{C_l} {=} \frac{C_l(\xi)-C_l}{C_l},
\label{eq:43}
\end{eqnarray}
where $C_l$ and $C_l(\xi)$ are the {coefficients} of angular power
spectrum {in a universe without the radiation-like solid and with
it} respectively. The values of $\Delta_\xi C_l / C_l$ are
compared with the cosmic variance for the {\it Planck} values of
cosmological parameters in fig. 2 and 3. The \linebreak

\begin{figure}[htb]
\centering
\includegraphics[scale=0.095]{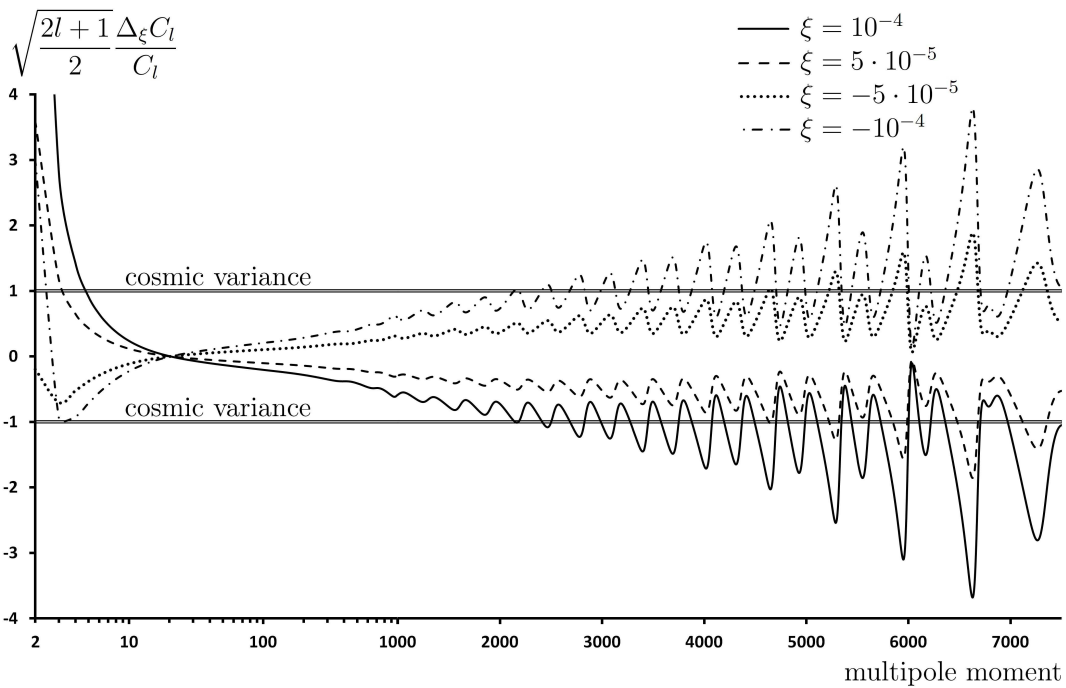}
\caption{{The ratio of the variance of the coefficients} of
angular power spectrum due to the presence of radiation-like solid
{to} the cosmic variance, depicted for given values of the shear
modulus parameter $\xi$. {On} the horizontal axis we have combined
logarithmic scale (for $l$ up to $1000$) with ordinary scale ($l$
over $1000$). The variance due to {the} presence of the
radiation-like solid surpasses the cosmic variance for very high
multipole moments as well as for very low ones if the integrated
Sachs--Wolfe effect is not considered. {With} the integrated
Sachs--Wolfe effect {included,} the effect of radiation-like solid
for low multipole moments is diminished, so that {the curves} get
under {the} cosmic variance for low multipole moments while not
changing for high ones.}
\end{figure}

\begin{figure}[htb]
\centering
\includegraphics[scale=0.125]{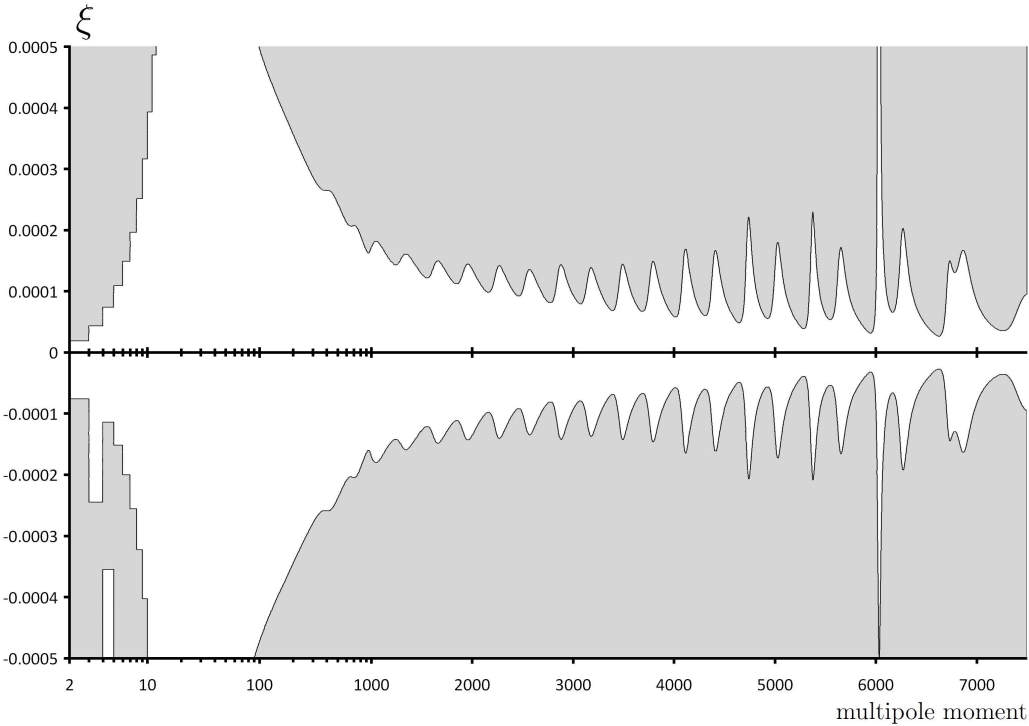}
\caption{{The values of the parameter $\xi$ for which the
variance} due to the presence of the radiation-like solid
{exceeds} the cosmic variance {(grey area)}. For the horizontal
axis we have combined logarithmic scale (for $l$ up to $1000$)
with ordinary scale ($l$ over $1000$).}
\end{figure}

\noindent integrated Sachs--Wolfe effect {was} not included {into
calculations} since its contribution for high multipole moments is
negligible. For {very} high multipole moments, {we have simplified
the calculation of the} integral (\ref{eq:41}) by using the
approximate} formula (9.72) in \cite{16}. Note that in order to
calculate {$C_l$} up to $l=7500$, one has to {know the} functions
$\alpha(\mathbf{k})$ and $\beta(\mathbf{k})$ {not just up to
$k\eta_* = 100$ as in fig. 1, but at least up to $k\eta_* \sim
150$.}

\vskip 2mm The coefficient $C_2(\xi)$ diverges due to the
contribution of long-wavelength part of the function $\tau_2(k)$
to the integral (\ref{eq:41}). {However,} because of finite
duration of inflatio{n} the wavelengths of quantum fluctuations
did not increase to infinity, {hence the integral (\ref{eq:41}) is
to be computed with nonzero lower limit $k_{min}$; then, no
divergence occurs}. The value of $k_{min}$ can be estimated as
$k_{min}\eta_0\sim e^{N_{min}-N}$, where $N$ is the number of
$e$-folds during inflation and $N_{min}$ is the minimal value of
$N$ needed to homogenize the universe {on} the scale of the
{present-day} Hubble radius. {We have used $k_{min}\eta_{*}=0.01$
($k_{min}\eta_0 \doteq 0.5$), which affects the angular power
spectrum only for the lowest multipole moments.}

\vskip 2mm {An explicit calculation shows that the enhancement of
viscosity due to the effects of polarization of photons, mentioned
after equation (\ref{eq:33}), while affecting the coefficients
$C_l(\xi)$ and $C_l$ of the angular power spectrum for high
multipole moments significantly, changes the ratio $\Delta_\xi
C_l/C_l$ only slightly. The reason is that for given multipole
moment the coefficients $C_l(\xi)$ and $C_l$ are changed by
similar factors. The same is true about the finite thickness
effect, and presumably also about the free-streaming of photons
and reionization. We thus believe that our results for high
multipole moments are not qualitatively different from the results
which one would obtain with the use of full Boltzmann equation, in
the sense that for $|\xi|\gtrsim 10^{-4}$ the absolute value of
$\sqrt{(2l+1)/2}\Delta_\xi C_l/C_l$ would still exceed 1 for
multipole moments starting from a value between 2000 and 3000
depending on $\xi$, equal approximately to that obtained in our
calculation.}

\vskip 2mm The presence of radiation-like solid can be confirmed
by observations only if $|\Delta_\xi C_l/C_l|$ ${>} |\Delta
C_l/C_l|$. {As seen from fig. 2 and 3, this holds for very high
$l$, even if $|\xi|$ is small enough so that inverse inequality is
valid for low $l$. To verify that the latter is true one can make
use of the result of \cite{15} that for low $l$ the integrated
Sachs--Wolfe effect suppresses the effect of radiation-like solid
on $C_l$ by a factor of about $50$}.

\vskip 1.5cm {\bf{\Large 5 Conclusion}}

\vskip 5mm We have applied perturbation theory in the proper-time
comoving gauge to a universe {which contains, in addition to}
radiation, baryonic matter and CDM, a radiation-like solid with
{the} pressure to energy density ratio $w=1/3$ and constant shear
modulus to energy density ratio $\xi$. The presence of such solid
does not change the evolution of the unperturbed universe.

\vskip 2mm In order to {obtain} results also for short-wavelength
perturbations, we had to extend the theory developed in \cite{15} and
{suppose that} the CDM and neutrinos {were} not coupled {to} the
baryon-radiation plasma. {This produced a} more complicated
{system of} equations; {however, as shown in the appendix, the
long-wave solution in both theories is the same. This provided us
with a simple check of the computed CMB anisotropies for very low
angular moments, since we could compare them directly with the
results of \cite{15}.}

\vskip 2mm We have calculated how a radiation-like solid
influences the CMB angular power spectrum {and found that its
presence has} a significant effect on the angular power spectrum
not only for very low multipole moments, as shown in \cite{15}, but
also for very high ones. {It turns out that the effect of the
solid is suppressed with the increasing multipole moment less
than} the error {of} the {coefficients} of angular power spectrum
{coming from the fact that we observe the universe just from one
place. The observational error caused by the dust in our Galaxy
decreases with multipole moment as well, therefore} future
observations of the CMB anisotropies with high enough resolution
could confirm or refute the presence of the radiation-like solid
in our universe with greater certainty than it is possible today.

\vskip 2mm For ${|\xi|}\sim10^{-4}$, the effect {is} observable
for $l\gtrsim3000$, which is the limit of {\it Planck} satellite
resolution. {At the same time,} according to \cite{15}, the
large-angle anisotropies are not affected significantly {for such
values of $\xi$ if one considers} also the integrated Sachs--Wolfe
effect. {We have} not included {this effect into our} wor{k}
because it {affects} the angular power spectrum significantly only
for low multipole moments; {however, if we take it into account we
find that} our model is in agreement with current observations for
{the} values of the shear modulus {parameter $\xi$ close enough to
zero,} while causing effects not predicted by {the} standard
theor{y} which are beyond the reach of current observations.

\vskip 2mm We have considered also negative $\xi$, which leads to
instability of vector perturbations \cite{14}, but the theory may still
be relevant since such perturbations are not created in standard
inflationary models. We have not included free streaming of
photons {into the theory}, which however should not change the
results qualitatively.

\vskip 4mm {{\it Acknowledgement.} The work was supported by the
grant VEGA 1/0985/16.}

\vskip 1.5cm {\bf{\Large {A} Long-wave solution}}
\setcounter{equation}{0}
\renewcommand{\theequation}{A-\arabic{equation}}

\vskip 5mm Equations of motion (\ref{eq:14}){--}(\ref{eq:19}) can
be solved {analytically} in the long-wavelength limit ($s\ll 1$).
In order to do so it is useful to rewrite {them in terms} of new
variable{s} $\Delta_E^d=\hat{\delta}^d-\mathcal{E}$,
$\Delta_B^d=\Theta-\tilde{\mathcal{B}}$,
$\Delta_E^\nu=\hat{\delta}^\nu-(4/3)\mathcal{E}$ and
$\Delta_B^\nu=\Sigma-\tilde{\mathcal{B}}$. {The resulting
equations are}
\begin{eqnarray}
\mathcal{E}'&=&-s^2\tilde{\mathcal{B}}-\alpha\tilde{\mathcal{H}}\epsilon
-\alpha_d\tilde{\mathcal{H}}\mathcal{D}_E^d-\alpha_\nu\tilde{\mathcal{H}}\mathcal{D}_E^\nu,
\label{eq:app1}\\
\tilde{\mathcal{B}}'&=&c_{s0}^2\epsilon+\Xi\mathcal{E}'+(\alpha-1)\tilde{\mathcal{H}}\tilde{\mathcal{B}}
+\alpha_d\tilde{\mathcal{H}}\Delta_B^d+\alpha_\nu\tilde{\mathcal{H}}\Delta_B^\nu+3\xi
c_{s0}^2\mathcal{E},
\label{eq:app2}\\
{\Delta_B^d}^\prime &=&-c_{s0}^2\epsilon-\tilde{\mathcal{H}}\Delta_B^d-3\xi
c_{s0}^2\mathcal{E},
\label{eq:app3}\\
{\Delta_E^d}^\prime &=&-s^2\Delta_B^d,
\label{eq:app4}\\
{\Delta_B^\nu}^\prime &=&-\left(c_{s0}^2-\frac{1}{3}\right)\epsilon+\frac{1}{4}\Delta_E^\nu-3\xi
c_{s0}^2\mathcal{E},
\label{eq:app5}\\
{\Delta_E^\nu}^\prime &=&-\frac{4}{3}s^2\Delta_B^\nu,
\label{eq:app6}
\end{eqnarray}
where $\alpha=3\rho_+/(2\rho)$,
$\epsilon=3\tilde{\mathcal{H}}\tilde{\mathcal{B}}+\mathcal{E}$,
$\mathcal{D}_E^d=3\tilde{\mathcal{H}}\Delta_B^d+\Delta_E^d$ and
$\mathcal{D}_E^\nu=3\tilde{\mathcal{H}}\Delta_B^\nu+(3/4)\Delta_E^\nu$.
{We will consider} this system of differential equations after the
{solidification time $\zeta_s$, when the parameter $\xi$ is
nonzero. Assuming} the solidification time to be close to zero,
$\zeta_s\ll 1$, the initial conditions are $\Delta_E^d=0$,
$\Delta_B^d=0$, $\Delta_E^\nu=0$, $\Delta_B^\nu=0$,
$\mathcal{E}=-(9/2)\Phi^{(0)}$ {and} $\mathcal{B}=0$. From now on
we set $\Phi^{(0)}=1$.

\vskip 2mm Our goal is to {determine} the potentials $\Phi$ and
$\Psi$ and {the} photon energy density contrast {$\delta_\gamma$},
which are needed to calculate the coefficients of the angular
power spectrum by using {the} integral (\ref{eq:41}). From the
first equation in (\ref{eq:23}) {it} follows
\begin{eqnarray}
\mathcal{E}'=s^2(\tilde{\mathcal{H}}^{-1}\Psi-\tilde{\mathcal{B}}),
\label{eq:app7}
\end{eqnarray}
therefor{e} in the long-wavelength limit {the function}
$\mathcal{E}$ can be approximated {by a constant}. The initial
condition {for $\mathcal{E}$ then yields}
\begin{eqnarray}
\mathcal{E}=-\frac{9}{2}. \label{eq:app8}
\end{eqnarray}
Similarly, from equations (\ref{eq:app4}) and (\ref{eq:app6}) {and
the corresponding initial conditions} we get
\begin{eqnarray}
\Delta_E^d = 0, \qquad \Delta_E^\nu = 0. \label{eq:app9}
\end{eqnarray}
{However, while the deviation of $\mathcal{E}$ from the value
(\ref{eq:app8}) is of order $s^2$, the deviations of $\Delta_E^d$
and $\Delta_E^\nu$ from zero are of order $s^4$.} We will use
these results in {the} following {considerations}.

\vskip 2mm {The} potential $\Psi$ can be rewritten with {the help}
of (\ref{eq:app1}) and the first equation in (\ref{eq:23}) as
\begin{eqnarray}
\Psi=\varphi-s^{-2}\tilde{\mathcal{H}}^2(\alpha_d\mathcal{D}_E^d
+\alpha_\nu\mathcal{D}_E^\nu),
\label{eq:app10}
\end{eqnarray}
where
$\varphi=-s^{-2}\tilde{\mathcal{H}}^2\alpha\epsilon$.
{Denote $X=a/a_{eq}=\zeta(\zeta+2)$.} Using equations of motion in
the zeroth and first order of the perturbation theory and
solutions (\ref{eq:app8}) and (\ref{eq:app9}){,} we can write the
{derivative} of $\varphi$ with respect to $\zeta$ as
\begin{eqnarray}
\varphi'=\left(\frac{f'}{f}+\frac{\hat{f}'}{\hat{f}}\right)\varphi+C+216\xi
s^{-2}f\hat{f}, \label{eq:app11}
\end{eqnarray}
where $f=(\zeta+1)X^{-3}$, $\hat{f}=(4+3X)/(4+3N_0x_bX)$, $N_0$
{being the} ratio of energy density of radiation including
neutrinos to photon energy density and $x_b$ {being the} ratio of
baryon energy density to {the} energy density of baryonic matter
and CDM, {and $C=\tilde{\mathcal{H}}^2\alpha \mathcal{B}$}. {The}
function $C$ can be {written} as
\begin{eqnarray}
C=\tilde{\mathcal{H}}^2\alpha\tilde{\mathcal{B}}_{\xi=0}+\mathcal{C}(\xi),
\label{eq:app12}
\end{eqnarray}
where $\tilde{\mathcal{B}}_{\xi=0}$ denotes {the} function
$\tilde{\mathcal{B}}$ {in a universe with no solid} (solution of
equations for perturbations {in which} the shear modulus parameter
$\xi$ is set to zero) and $\mathcal{C}$ is a function of {first}
order in $\xi$. {Obviously, the} function $\mathcal{C}$ is much
smaller than the third term in (\ref{eq:app11}) in the
long-wavelength limit, {so that the} function $C$ can be computed
assuming that $\xi$ is zero.

\vskip 2mm {With $\cal E$} given by (\ref{eq:app8}) and {$s \ll
1$,} equation (\ref{eq:app1}) reduces to
$\alpha\epsilon+\alpha_d\mathcal{D}_E^d+
\alpha_\nu\mathcal{D}_E^\nu=0$. {Furthermore,} by inserting
{from} (\ref{eq:app9}) into {the} definitions {of}
$\mathcal{D}_E^d$ and $\mathcal{D}_E^\nu$ we get
\begin{eqnarray}
\mathcal{D}_E^d=3\tilde{\mathcal{H}}\Delta_B^d, \qquad
\mathcal{D}_E^\nu=3\tilde{\mathcal{H}}\Delta_B^\nu.
\label{eq:app13}
\end{eqnarray}
Hence, if $\xi$ is zero, $\epsilon$ can be expressed as
{a} linear combination of $\Delta_B^d$ and $\Delta_B^\nu$ (with
{the coefficients depending on the} time $\zeta$) and, therefore,
equations (\ref{eq:app3}) and (\ref{eq:app5}) can be reduced to
{a} system of linear differential equations for $\Delta_B^d$ and
$\Delta_B^\nu$. The only solution of this system {satisfying
the} initial conditions is $\Delta_B^d=0$ and $\Delta_B^\nu=0$,
which implies that $\epsilon$ must be zero {as well}.
Therefore, using {the} definition of $\epsilon$ and
(\ref{eq:app8}){,} we get
{$\tilde{\mathcal{B}}$}${}_{\xi=0}=3/(2\tilde{\mathcal{H}})${,}
and {by inserting this into the expression for the} function $C$
{we find that $C$} can be approximated as
\begin{eqnarray}
C=\frac{3}{2}\tilde{\mathcal{H}}\alpha=\frac{3}{2}{\frac{4+3X}{X(\zeta+1)}}.
\label{eq:app14}
\end{eqnarray}

\vskip 2mm {After determining the unknown function on the right
hand side of the} differential equation (\ref{eq:app11}), we can
find $\varphi$ by solving this equation. {With the} substitution
$\varphi=f\hat{f}F$, equation (\ref{eq:app11}) {yields}
\begin{eqnarray}
F'=\frac{C}{f\hat{f}}+216\xi s^{-2},
\label{eq:app15}
\end{eqnarray}
and {by a straightforward integration we obtain}
\begin{eqnarray}
F=6{\zeta^3\left[\varepsilon\zeta\left(\frac{1}{5}\zeta+1\right)+\frac{1}{3}
+\varepsilon +(1-\varepsilon)\frac 1{\zeta+1}\right]}+216\xi
s^{-2} \zeta, \label{eq:app16}
\end{eqnarray}
where $\varepsilon=(3/4)N_0 x_b$. {In order that the contribution
of the solid to the perturbations is comparable to or less than
the contribution of the remaining components of matter, we require
that $\xi s^{-2} \lesssim 1$.}

\vskip 2mm {To compute $\Psi$ we need to know, in addition to the
function $\phi$, the functions $\mathcal{D}_E^d$ and
$\hat{\tilde\Delta}_E$ up to the order $s^2$. However, as
mentioned before, the functions $\Delta_E^d$ and $\tilde\Delta_E^d$
are of order $s^4$; thus, it suffices to know the functions
$\Delta_B^d$ and $\tilde\Delta_B^d$ up to the order $s^2$. These
functions are given by equations (\ref{eq:app3}) and
(\ref{eq:app5}), which can be rewritten as}
\begin{eqnarray}
{\Delta_B^d}^\prime=s^2\frac{D}{X}f\hat{f}F
-\tilde{\mathcal{H}}\Delta_B^d+\frac{27}{2}\xi c_{s0}^2,
\label{eq:app17}
\end{eqnarray}
and
\begin{eqnarray}
{\Delta_B^\nu}^\prime=-s^2\varepsilon Df\hat{f}F+\frac{27}{2}\xi
c_{s0}^2, \label{eq:app18}
\end{eqnarray}
where $D=Xc_{s0}^2/(\tilde{\mathcal{H}}^2\alpha)={(3/2)
X^4/(4+3X)^2}$ and the sound speed {squared can be written} as a
function of {time as} $c_{s0}^2={1/[3(1+\varepsilon X)]}${. The
latter equation integrates immediately and the former equation can
be integrated as well after passing from $\Delta_B^d$ to
$X\Delta_B^d$.}

\vskip 2mm {For the potential $\Psi$ we have}
\begin{eqnarray}
\Psi=f\hat{f}F-s^{-2}{(E_d\Delta_B^d+E_\nu\Delta_B^\nu)},
\label{eq:app19}
\end{eqnarray}
where $E_d=3\tilde{\mathcal{H}}^3\alpha_d=36x_d fX$ and
$E_\nu=3\tilde{\mathcal{H}}^3\alpha_\nu=48\nu f${,} with CDM
energy density to energy density of CDM and baryonic matter ratio
denoted {by} $x_d$ and neutrino energy density to {the} energy
density of all kinds of radiation ratio denoted {by} $\nu$,
{$x_d=1-x_b$ and $\nu=1-N_0^{-1}$}. Inserting {for $F$ the
expression (\ref{eq:app16}) and for $\Delta_B^d$ and
$\tilde\Delta_B^d$ the expressions obtained by integrating equations
(\ref{eq:app17}) and (\ref{eq:app18}), after some algebra we
arrive at the formula derived in \cite{15} for a universe in which all
kinds of matter are coupled to each other. The formula reads}
\begin{eqnarray}
\Psi={\frac32f\zeta^3\left(\frac35\zeta^2+3\zeta+
\frac{13}3+\frac1{\zeta+1}\right)}+216{\tilde\xi} s^{-2}f\zeta.
\label{eq:app20}
\end{eqnarray}
where $\tilde \xi = \mu/\rho^{sr}$, $\rho^{sr} = N_0
\rho^{s\gamma}$ being the total energy density of the solid and
radiation (photons and neutrinos).

\vskip 2mm {Having determined $\Psi$, we can compute the Newtonian
potential $\Phi$ from} the second equation in (\ref{eq:23}). {We
obtain immediately the same expression as in \cite{15},}
\begin{eqnarray}
\Phi=\Psi-108{\tilde\xi}s^{-2}\frac1{X^2}. \label{eq:app21}
\end{eqnarray}
{Finally, from equation (\ref{eq:25}) we find that the} photon
energy density contrast is $\delta_\gamma = 4\Psi-6$. As a
result, the function $\alpha({\bf k})$ can be written in terms of
$\Phi$ and $\Psi$ as $\alpha({\bf k})= \Phi+\Psi-3/2$.

\vskip 2mm In {the} calculation of {the coefficients of angular
power spectrum} we need {the value of the function $\alpha({\bf
k})$} at the time of recombination $\zeta=\zeta_r$. For
cosmological parameters {which are} obtained {from the data
assembled by the} satellite {{\it Planck} we have $\zeta_r =
1.062$, which yields $\alpha_r({\bf k})=0.371+17.3\tilde\xi
s^{-2}$.}

\vskip 2mm To find {the} contribution of Silk damping to the
temperature fluctuations we need to extend the expression for
$\Phi$ by {the} viscosity term which has been omitted in {the}
second equation (\ref{eq:23}). {(Viscosity term in the equations
for perturbations was kept, since it plays an important role for
wavenumbers which are not small.)} {Viscosity contributes to
$\Phi$ and $\Psi$ by terms of order $k^0$ or $\xi k^{-2}$} so that
it is {in principle} relevant in the long-wavelength limit{,
however}, because the {viscosity coefficient} is small, the Silk
damping can be neglected in the considered limit. Since the
function $k\beta(\mathbf{k})$ is zero in this limit, the Doppler
effect can be neglected as well. {On the other hand}, contribution
of integrated Sachs--Wolfe effect cannot be neglected for small
multipole moments. To calculate it one can use relations (64) in
\cite{15}{, with additional terms originating in the decoupling of
CDM and neutrinos}.

\end{document}